\documentclass{iau} 
\usepackage{graphicx}

\title[JD 11.~~Extragalactic Supergiants] %% give here short title %%
{Extragalactic Supergiants}

\author[Urbaneja \& Kudritzki]   %% give here short author list %%
{Miguel A. Urbaneja$^1$
 \and Rolf P. Kudritzki$^2$}

\affiliation{$^1$Institut f\"ur Astro- und Teilchen-Physik, Universit\"at Innsbruck, \\ Technikerstr. 25/8, A-6020,
, Innsbruck, Austria \\ email: {\tt Miguel.Urbaneja-Perez@uibk.ac.at} \\[\affilskip]
$^2$Institute for Astronomy, University of Hawaii, \\ 2680 Woodlawn Drive, 
Honolulu HI 96822, USA \\email: {\tt kud@ifa.hawaii.edu}
}

\pubyear{2016}
\volume{329}  %% insert here IAU Symposium No.
\jname{The lives and death-throes of massive stars}
\editors{A.C. Editor, B.D. Editor \& C.E. Editor, eds.}
\begin{document}

\maketitle

\begin{abstract}
Blue supergiant stars of B and A spectral types are amongst the visually brightest non-transient astronomical objects. Their intrinsic brightness
makes it possible to obtain high quality optical spectra of these objects in distant galaxies, enabling the study not only of these stars
in different environments, but also to use them as tools to probe their host galaxies.  Quantitative analysis of their optical spectra provide tight 
constraints on their evolution in a wide range of metallicities, as well as on the present-day chemical composition, extinction laws and 
distances to their host galaxies. We review in this contribution recent results in this field. 
%% \log\,g_{\!\mbox{\scriptsize \,\sc f}}
%%% $T_{\!\mbox{\scriptsize \em eff}}^4$

\end{abstract}

\firstsection % if your document starts with a section,
\section{Introduction}
Blue supergiant stars with B and A spectral types are amongst the visually brightest non-transient astronomical objects. They are the descendants of stars
born in the mass range between 15 to 40 M$_\odot$, in a short-lived evolutionary phase, after the end of the H-core burning phase. %The short duration of this 
%phase along with the reduced rates (compared to their Main Sequence counterparts) at which they loose mass via stellar winds, coupled with an almost constant 
%luminosity, produces \ldots 
Their intrinsic brightness
makes it possible to obtain high quality optical spectra of these objects in distant galaxies, enabling the study not only of these stars
in different environments, but also to use them as tools to probe their host galaxies. Their optical spectra are rich in metal absorption lines from many different elements 
(among others, C, N, O, Mg, Si, S, Ti, Fe). As young objects, with ages of a few Myr, they provide key probes of the current chemical composition of the interstellar medium.  
Quantitative interpretation of these optical spectra by means of sophisticated model atmosphere/line formation codes, coupled with efficient analysis methods, allowed in 
recent years the study of 
individual supergiants in galaxies well beyond the realm of the Local Group, providing information not only on the characteristics of the stars, but 
also opening a window to investigate properties of their host galaxies, such as reddening, extinction laws, chemical composition and distances.

Whilst studies of these objects within the Local Group have been conducted for a while (see \cite[Kudritzki et al. 2008a]{kud2008iau} and references therein), the jump to more distant galaxies is relatively recent, the main reasons being the availability of very efficient multi-object spectrographs attached to the generation of 10m class telescopes, in combination with
advanced radiative transfer models that properly treat the physics describing the atmospheres of these objects (intense radiation fields propagating in low density environments) and the development of novel analysis techniques. This last is a crucial point, since spectral resolution has to be sacrificed in order to achieve very good signal-to-noise ratios for faint objects. Following on the first ideas presented in \cite[Kudritzki et al. (1995)]{kud95}, the seminal papers by \cite[Kudritzki et al. (2008b)]{kud2008} for the case of 
mid B to A-type supergiants, and by \cite[Urbaneja et al. (2005)]{urb2005} for the case of early B-type supergiants, represent the cornerstones for the quantitative analysis of low resolution ($\lambda/\Delta\lambda\sim1000$) optical spectra of these objects. Besides obtaining information on the stellar parameters and chemical composition of the stars in NGC\,300, a mid size disk galaxy located at 1.9 Mpc, these works allowed the first detailed study of the stellar metallicity gradient in an external galaxy. More importantly, they confirmed the existence of the so-called \textquotedblleft Flux-weighted Gravity--Luminosity Relationship\textquotedblright~initially introduced by 
\cite[Kudritzki et al. (2003)]{kud2003}. \\

Blue supergiants  have long been recognised as potentially distance indicators in external galaxies (see for example \cite[Hubble 1936]{hubble1936},  
\cite[Tully \& Wolff 1984]{tully1984}).
However, the harsh conditions in their atmospheres required a significant amount of development in the non-LTE model atmosphere techniques for them to be used 
reliably (\cite[Przybilla et al. 2006]{przybilla2006}). The work by \cite[Kudritzki et al. (2008b)]{kud2008}  
based on quantitatively analyzing high signal-to-noise optical spectra to determine stellar parameters
such as effective temperature $T_{\!\mbox{\scriptsize \em eff}}$, gravity $\log g$~and metallicity with unprecedented accuracy and reliability, led to the 
detection of a tight relationship between stellar absolute bolometric magnitude M$_{\rm bol}$ and the flux-weighted gravity $g/T_{\!\mbox{\scriptsize \em eff}}^4$ 
($\equiv$ $g_{\!\scriptsize \,\sc f}$), the \textquotedblleft Flux-weighted Gravity - Luminosity Relationship (FGLR)\textquotedblright, of the form
\begin{equation}
% M_{\rm bol}\,=\,a (\mbox{\log\,g_{\!\mbox{\scriptsize \,\sc f}}}\,-\,1.5)\,+\,b, 
 M_{\rm bol}\,=\,a (\log\,g_{\!\scriptsize \,\sc f}\,-\,1.5)\,+\,b, 
\end{equation}
A very basic back-of-the-envelope calculation confirms that this simple form of the FGLR 
can be understood as the result of (single) massive star evolution beyond the main sequence, in the crossing to the red supergiant phase at constant luminosity and 
constant mass. Under these simplifying assumptions, the quantity $g/T_{\!\mbox{\scriptsize \em eff}}^4$ remains constant 
during the evolution accross the Hertzsprung-Russell Diagram and the FGLR is then the result of a simple relationship between stellar luminosity 
$L$ and stellar mass $M$, $L \sim M^x$ (see \cite[Kudritzki et al. 2008b]{kud2008} , for further details). Even though the luminosity is not exactly constant during 
the evolution (also the exponent $x$ decreases with increasing stellar mass), the detailed study by \cite[Meynet et al. (2015)]{meynet2015} combining state-of-the-art stellar 
evolution calculations (for single stars) with population synthesis has recently shown that the FGLR still holds. 

\section{The case for alternative metallicity indicators}
The spatial distribution of chemical elements in galaxies represents a key tool to understand how the chemical evolution proceeds in the universe. In the case of star-forming galaxies, the classical procedure has been to use oxygen abundances derived from emission lines present in the spectra of H~{\sc ii} regions for studies of abundance gradients in individual galaxies, and more recently the existence of a relationship between the mass of the galaxy and their {\em metal} content 
(\cite[Tremonti et al. 2004]{tremonti2004}). Gas phase oxygen abundances have been also widely used to interpret results obtained from studies of populations of massive stars, such as the dependence of Wolf-Rayet types or the dependence of the number of Blue-to-Red supergiants with the {\em metallicity} of the environment (see \cite[Massey 2003]{massey2003} and references therein), as well as for studies of the {\em metallicity} dependence of the Period--Luminosity relationship of Cepheid stars, with far reaching implications for the distance scale of the universe (\cite[Freedman \& Madore 2010]{freedman2010}, \cite[Bono et al. 2010]{bono2010}). \\

All these studies relied on the analysis of the emission spectrum of
H~{\sc ii} regions because it is very easy to collect high quality spectra for these objects even in very distant galaxies, and under the key assumption that their physics was quite simple. In recent years however, it has become very clear that these objects are not as simple as originally thought, and that their physics has yet to be properly understood (\cite[Stasi{\'n}ska 2008]{stasinska2008}). Among others, the dependence of the derived oxygen abundances upon the choice of calibration used to interpret the ratio of some of the most prominent forbidden lines (the so-called statistical or strong-line methods, see for example \cite[Bresolin et al. 2009]{bresolin2009}), the large discrepancies between abundances based on the derivation of the electron temperature of the gas and the ones obtained from statistical methods (both applied to the same set of spectra) or the systematic difference between abundances based on collisionally excited lines, CELs, and on recombination lines, RLs, when observed in the same object (termed in the literature as the \textquotedblleft Abundance Discrepancy Factor\textquotedblright, ADF, see for example \cite[Toribio San Cipriano et al. 2016]{toribio2016}), are sources of concern when interpreting any observed property in terms of the {\em metal} content of the host galaxy. Furthermore, the evolution of massive stars is at the very least the result of the interplay between rotation and mass-loss (\cite[Meynet \& Maeder 2005]{meynet2005}). Whilst light elements, such as oxygen, will play a role in the acceleration of the stellar wind in the outer layers of the atmosphere, hence having a significant impact on the wind terminal velocity, the mass-loss rate is set at the base of the wind, where the radiative acceleration is provided by a myriad of iron lines (\cite[Vink et al. 2000]{vink2000}, \cite[Puls et al. 2000]{puls2000}). Thus, any proper interpretation of the wind properties of massive stars, such as the Wind Momentum--Luminosity Relationship (WLR, \cite[Kudritzki \& Puls 2000]{kud2000}) and its dependence with {\em metallicity} requires the knowledge of the iron content. Since the $\alpha$-to-iron ratio (oxygen being an $\alpha$-element) in any stellar system will depend not only on the original composition, but also on how the system has evolved in time (oxygen in mainly produced when massive stars die, whilst iron is mainly produced by less massive stars), there is a priori no reason for the ratio observed in the solar vicinity to apply universally (as sometimes is assumed). 

\begin{figure}[!h]
% \vspace*{-2.0 cm}
\begin{center}
 \includegraphics[width=5.0in]{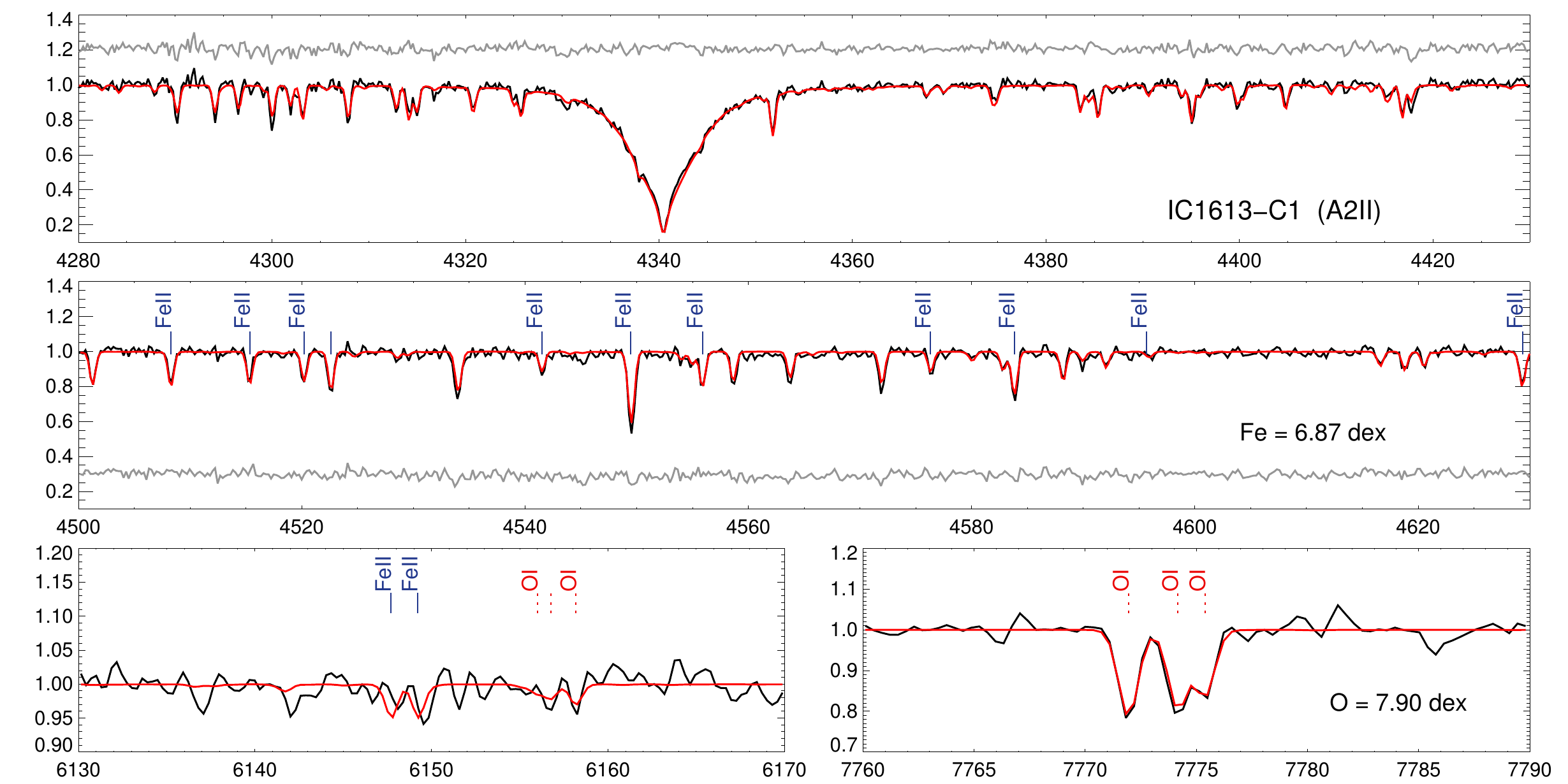} 
% \vspace*{-1.0 cm}
 \caption{Observed X-Shooter (R$\sim$9900) spectrum of an A-type supergiant in IC\,1613 (black) with the final tailored model (red). These analyses confirm the low oxygen content as well as the sub-solar O/Fe ratio (Urbaneja et al., in prep). }
   \label{fig1}
\end{center}
\end{figure}

The case of the {\em too strong} winds in the putative low {\em metallicity} Local Group galaxy IC\,1613 illustrates this point. Recent studies of massive stars in this galaxy 
(among others \cite[Tramper et al. 2011]{tramper2011}, \cite[Herrero et al. 2012]{herrero2012}; see also the contribution by Miriam Garcia in these proceedings) observed an 
apparent over strength of the stellar winds when comparing the WLR of stars in IC1613 with Galactic and LMC and SMC counterparts, with respect to predictions provided 
from the  theory of Radiatively Driven Winds (\cite[Kudritzki \& Puls 2000]{kud2000}). However, the interpretation of these results were based on the oxygen content of this
galaxy. Here we point out that previous studies of blue type supergiants in this galaxy (\cite[Bresolin et al. 2007]{bresolin2007})
confirmed the low oxygen abundances derived from H~{\sc ii} regions, about 1/10th of the solar neighbourhood value. However, upon a closer
inspection, it turns out that the O/Fe ratio in this galaxy is sub-solar, with the current Fe content of IC\,1613 being somewhere
closer to the SMC value (\cite[Tautvai{\v s}ien{\.e} et al. 2007]{taut2007}). A recent high resolution study of A-type supergiants in this galaxy (Urbaneja et al.
in prep., see Fig.~\ref{fig1}) confirms the oxygen abundances from H~{\sc ii} regions and B-type supergiants as well as sub-solar O/Fe
ratio.  When the WLR results are interpreted in terms of the iron content of the different galaxies, the results fully support the theoretical predictions.\\
      
As indicated above, oxygen abundances derived from emission line spectra of ionised gas are
affected by a number of systematic problems, not yet understood. One way to tackle these issues, at least in part, is through
the comparison of several different abundance indicators. Whilst it is true that spectroscopy of individual stars can not reach the far
distances covered with nebular studies, it is possible to study nearby systems, in order to investigate which, if any, of the nebular diagnostics
is more reliable, and hence should be preferred over the others. A number of nearby galaxies have been studied in recent years with this purpose in mind. Comparative 
studies of H~{\sc ii} regions and blue supergiants seem to be yielding some fruitful results. Very recently, \cite[Bresolin et al. (2016)]{bresolin2016} presented a comparison of abundances
in different galaxies (both in terms of the oxygen content and the mass of the galaxies). While still not definitive, it seems that CEL-based abundances (when the electron temperature of the gas is known) are fully consistent with stellar abundances for oxygen abundances below the solar neighbourhood content, whilst RL-based abundances seem to be systematically higher. On the other hand, the situation reverts for oxygen abundances above the solar neighbourhood value. This result needs to be confirmed with a larger sample of objects in this second regime. However, upon confirmation, this would help to constrain the physical origin of the ADF problem.

\section{A stellar based Mass--Metallicity relationship for nearby galaxies}
Based on the analysis of more than 40000 star forming galaxies, \cite[Tremonti et al. (2004)]{tremonti2004} showed that there seems to be a relationship between the amount of oxygen in a 
galaxy and its mass. This \textquotedblleft Mass--Metallicity\textquotedblright relationship of star-forming galaxies represents a very powerful tool to investigate the chemical 
evolution of the universe, by for example comparing the relationship observed at different redshifts (\cite[Zahid et al. 2014]{zahid2014}). As is customary, oxygen abundances derived by means 
of a specific strong line method were used in this work as proxy for metallicities. However, the placement of our nearest companions (the Magellanic Clouds) and the Milky 
Way in this relationship, when using what is known about the oxygen content of their stars, results in the oddity that these three galaxies are 3$\sigma$ outliers in this relationship. This 
is at least in part related to the choice of the strong line indicator used to derive gas phase oxygen abundances.
In a later work, \cite[Kewley \& Ellison (2008)]{kewley2008} illustrated how not only the absolute abundance scale 
but also the shape of the relationship would depend upon the choice of the 
calibration. To overcome the potential problems that could appear when comparing results obtained by different authors using different strong line diagnostics, these authors 
provided relative calibrations that allow transformation between the different diagnostics. Thus, everything would potentially be fine when working in relative terms. However, there are applications for which an {\em absolute} scale is required, like for example to investigate the metallicity dependence of the PL relationship of Cepheid stars. 

\begin{figure}[!h]
% \vspace*{-2.0 cm}
\begin{center}
 \includegraphics[width=3.1in]{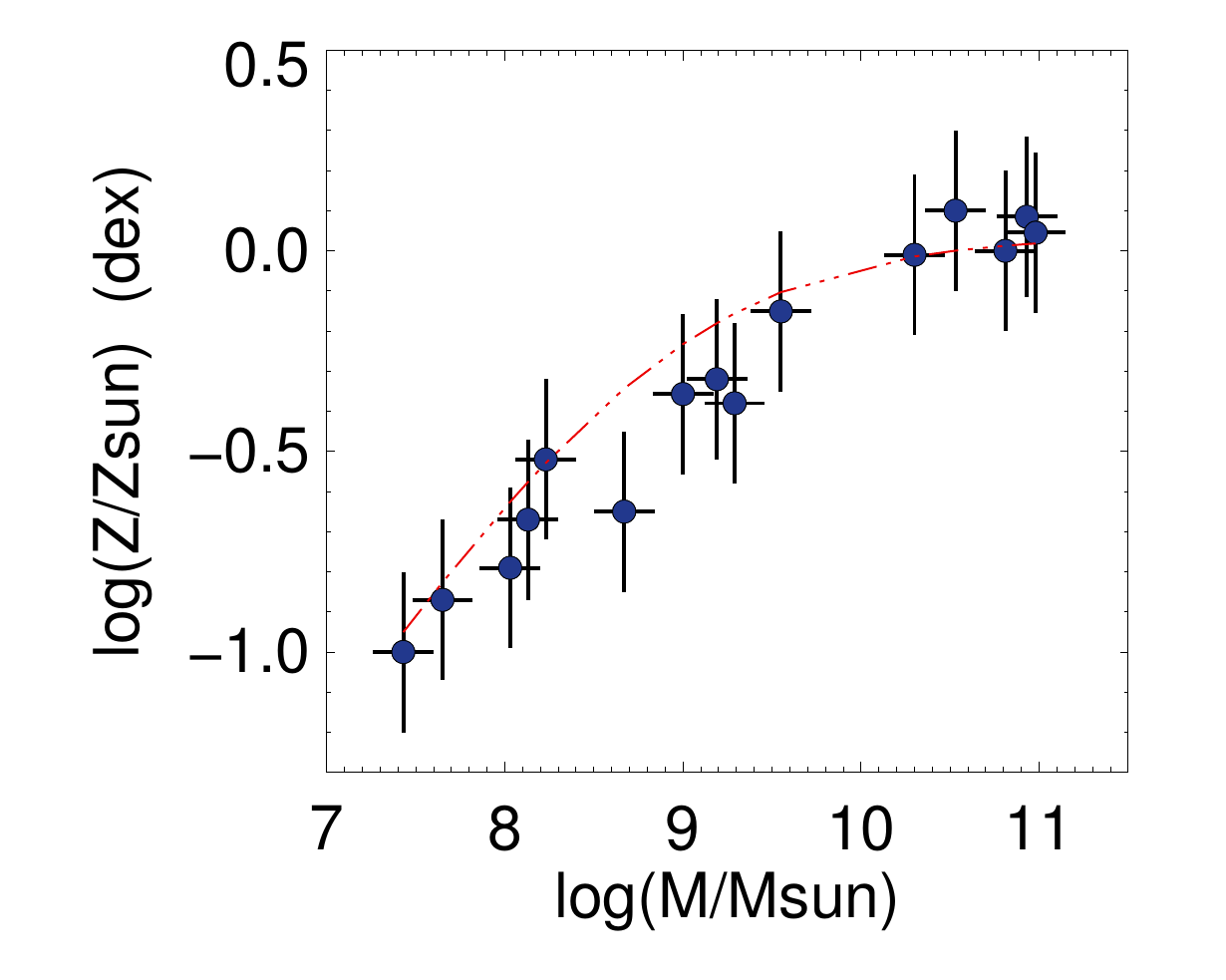} 
% \vspace*{-1.0 cm}
 \caption{The \textquotedblleft stellar\textquotedblright Mass--Metallicity relationship for nearby star-forming galaxies. Metallicities are obtained from the quantitative spectroscopy of individual blue supergiant stars in these galaxies.}
   \label{fig3}
\end{center}
\end{figure}

It is possible to create a similar mass--metallicity relationship for nearby galaxies based only on stellar abundances derived from individual blue supergiant stars, albeit with a much smaller number of galaxies. This is shown in Fig.~\ref{fig3}, where abundances derived from individual blue supergiant stars and stellar masses for the galaxies have been collected from the literature. The red line shown in the figure is not a fit to the data, but just a simple zero point adjustment of the Mass--Metallicity relationship obtained by \cite[Andrews \& Martini (2013)]{andrews2013}, based on staked spectra (using mass bins) of the same galaxy sample studied by  \cite[Tremonti et al. (2004)]{tremonti2004} and \cite[Kewley \& Ellison (2008)]{kewley2008}.

\section{A Spectroscopic distance indicator: the FGLR}
Along with metallicities, knowledge of distances to galaxies represent a key ingredient in our understanding of the universe. The primary stellar distance indicator, the
Period--Luminosity relationship of Cepheid stars, suffers from two major problems: extinction and metallicity dependence, both of which are difficult to estimate individually 
with the stringent precision required to reduce the uncertainties in measuring the Hubble Constant $H_0$ (with this being ultimately related to constraining the equation of 
state of 
Dark Energy, \cite[Riess et al. 2016 ]{riess2016}). Quantitative spectroscopy of Cepheids beyond the Magellanic Clouds is currently not feasible, because of the intrinsic brightness of these objects 
(about 3 orders of magnitude fainter than blue supergiant stars). Hence, in order to gauge the effect 
that the metallicity may or may not have in the PL relationship, alternative metallicity indicators are used as proxies. Mainly, gas phase oxygen abundances derived
from H~{\sc ii} regions. Therefore any interpretation in terms of {\em metallicity} is severely hampered by the unknown systematics discussed above. The second source of 
concern is related to internal extinction in the host galaxies. Whilst it is true that there are ways to try to circumvent this problems (by using 
\textquotedblleft reddening-free\textquotedblright magnitudes, or moving to the IR domain), there is still no clear consensus on the accuracy of these 
methods. Moreover, the lack of information 
concerning the real form of the extinction law is always present (see \cite[Kudritzki \& Urbaneja 2012a]{kud2012} for a detailed discussion).

Blue supergiant stars can also contribute in this field. As young objects, they are closely related to Cepheids, and hence expected to present similar extinction values (note 
that we are referring here exclusively to B- and A-type supergiant stars). First, for each individual blue supergiant, information regarding its metal content is directly obtained 
from the analysis of the 
optical spectrum. At the same time, reddening as well as information about the form of the extinction curve can be obtained individually for each star, by combining the 
intrinsic colours predicted by the model atmosphere corresponding to the physical parameters describing the observed spectrum with precise multi-band photometric 
observations. This method has been for example used in several galaxies (NGC\,300--\cite[Kudritzki et al. 2008]{kud2008}, M33--\cite[U et al. 2009]{u2009}, 
M81--\cite[Kudritzki et al. 2012b]{kud2012b}) to investigate the differences between the observed E(B-V) values derived from blue supergiants with the foreground values customarily applied in the Cepheid's literature.  The basic conclusion from these comparisons is that the concept of applying a constant foreground reddening correction for Cepheid stars is a dangerous notion (\cite[Kudritzki \& Urbaneja 2012a]{kud2012}).

Besides these kind of comparisons, blue supergiants can be used to derive distances to nearby galaxies by means of the FGLR. As discussed in the introduction, there is a 
tight observed correlation between the absolute bolometric magnitude and the flux-weighted gravity of blue supergiant stars (see eq. 1.1). Once this relationship is 
calibrated in absolute terms with stars in a galaxy with a well know distance, the \textquotedblleft apparent\textquotedblright~FGLR (based on apparent bolometric 
magnitudes) of stars in any other galaxy can be compared with the FGLR calibrator, and the difference in the intersects will directly provide the relative distance between 
both galaxies. FGLR based distances have been derived for a number of nearby galaxies, using NGC\,300 as calibrator: 
WLM--\cite[Urbaneja et al. 2008]{urbaneja2008}, M33--\cite[U et al. 2009]{u2009}, M81--\cite[Kudritzki et al. 2012b]{kud2012b}, 
NGC\,3109--\cite[Hosek et al. 2014]{hosek2014}, 
NGC\,3621--\cite[Kudritzki et al. 2014]{kud2014}, NGC\,55 --\cite[Kudritzki et al. 2016]{kud2016} and M83--\cite[Bresolin et al. 2016]{bresolin2016}. Excluding the yet to be understood large discrepancy for the case of M33 (\cite[U et al. 2009]{u2009}, see also \cite[Bonanos et al. 2006]{bonanos2006}), the FGLR derived distances are in good agreement with distances derived by using multi-band photometry (covering the optical regime as well as the J and K bands) of Cepheid stars. 

\begin{figure}[!h]
% \vspace*{-2.0 cm}
\begin{center}
 \includegraphics[width=4.5in]{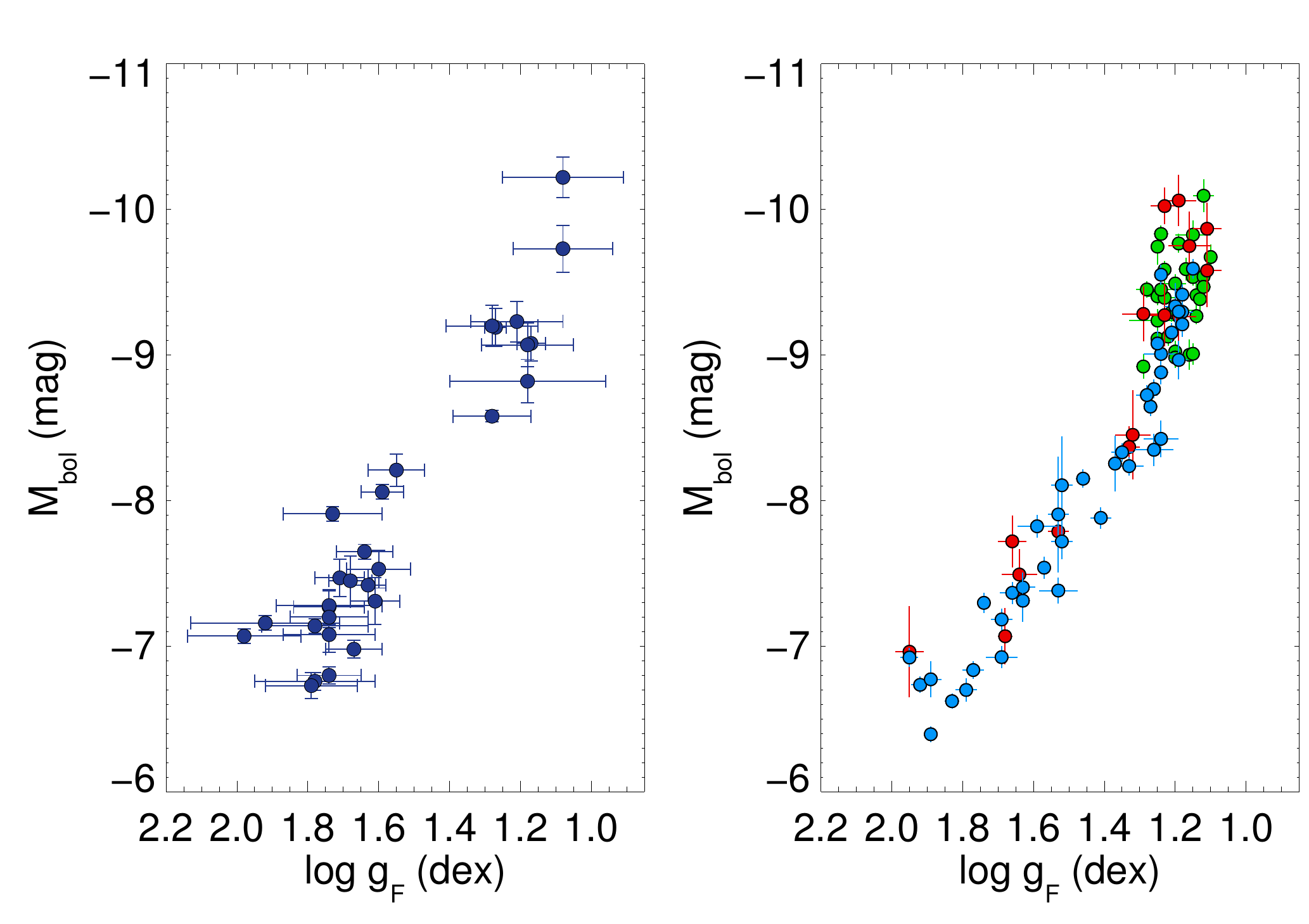} 
% \vspace*{-1.0 cm}
 \caption{The FGLR of blue supergiant stars. Left--current FGLR
 calibrator: stars in NGC\,300 (adapted from \cite[Kudritzki et
 al. 2008]{kud2008}). Right--stars in the LMC (Urbaneja et al., in prep).}
   \label{fig2}
\end{center}
\end{figure}

\section{Blue supergiants in era of the ELTs.}
The future introduction of the Extremely Large Telescopes (E-ELT, TMT, GMT) will significantly enlarge the volume of the universe in which quantitative spectroscopy of 
individual stars would be feasible. Whilst these new towering giants are being designed primary to work in the IR domain, some capabilities in the optical will still be present, 
enabling the use of individual blue supergiant stars for studies of their host galaxies well beyond the limits of current facilities. But it would be in the IR domain, thanks to the 
extreme AO capabilities, where these telescopes will provide the maximum efficiency. Most likely, B- to early A-type supergiants will have a diminishing impact in the 
extragalactic field. However, these new facilities will strengthen the relevance of red supergiants (see Ben Davies' contribution in these proceedings) and 
some newcomers to the extragalactic playground: yellow supergiants, with spectral types from late A to early G. 

\begin{figure}[!h]
% \vspace*{-2.0 cm}
\begin{center}
 \includegraphics[width=3.1in,angle=270]{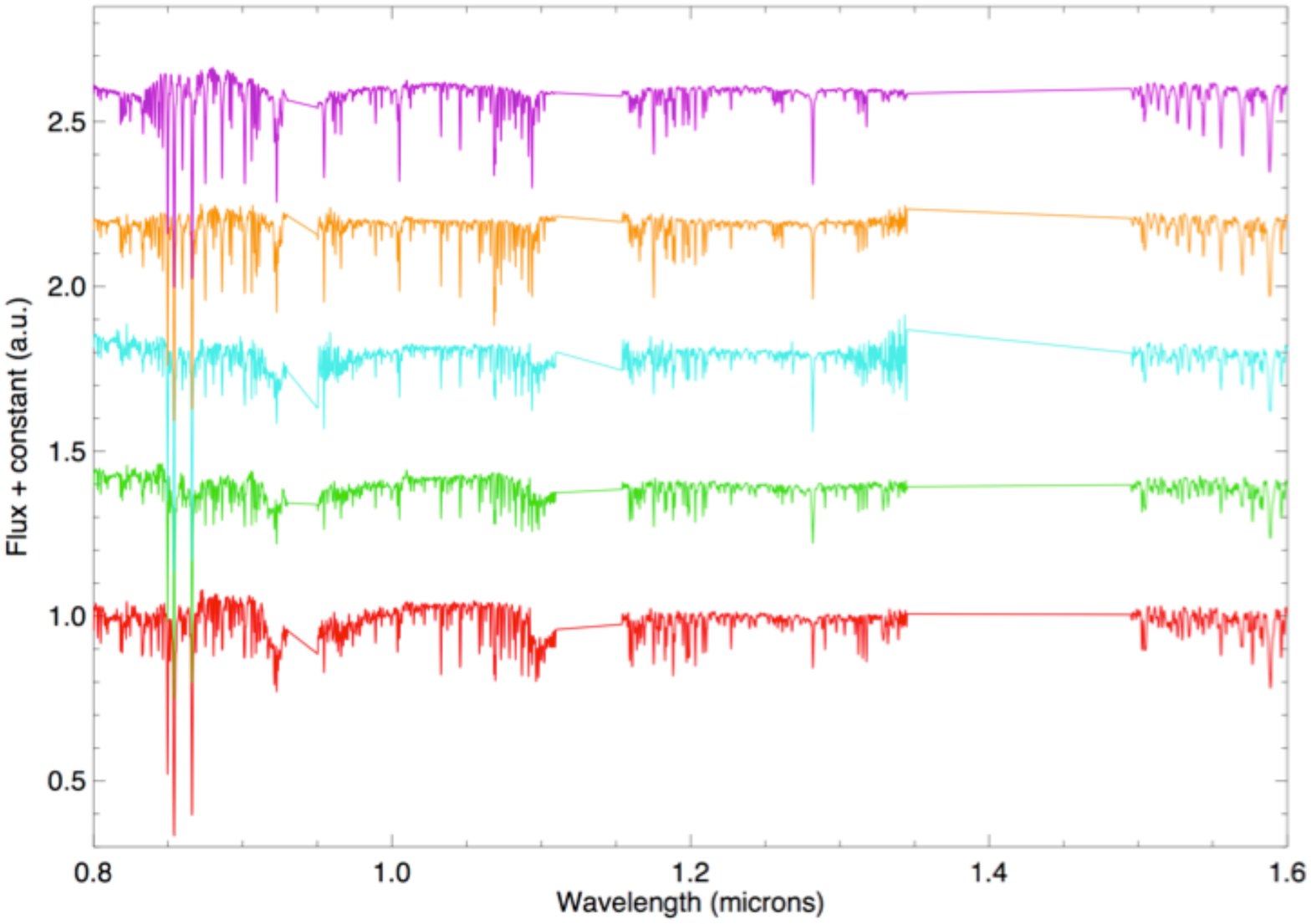} 
% \vspace*{-1.0 cm}
 \caption{IR spectra of Galactic F-type supergiants from the IRTF Spectral library (\cite[Rayner et al. 2009]{rayner2009}, R$\sim$2000) in the 0.8--1.6 $\mu$m spectral window. These illustrate the potential of yellow supergiants as metallicity indicators in external galaxies.}
 \label{fig4}
\end{center}
\end{figure}

As part of the first-light instrument suite for ESO's E-ELT, the Multi-Adaptive Optics Imaging Camera for Deep Observations (MICADO, 
\cite[Davies et al. 2016]{davies2016})  is being design to provide 
spectroscopic capabilities with very high spatial resolution in the 0.8--2.4 $\mu$m wavelength range, split into two settings, zYJ and HK, 
at a spectral resolution of R$\sim$8000. Fig.~\ref{fig4} displays a sequence of real IR spectra of F-type supergiant stars from the IRTF Spectral Library 
(\cite[Rayner et al. 2009]{rayner2009}), in the 0.8--1.6 $\mu$m at a lower spectral resolution R$\sim$2000. Many individually resolved atomic lines 
from several species are contained in this region. Thus, in principle, the same techniques that have been successfully employed in the optical for 
B- and A-type supergiants could potentially be applied for the quantitative 
analysis of the IR spectrum of these objects, provided that the models atmospheres (and corresponding model atoms) are fully developed for their application in this 
scarcely explored regime.

{\underline{Acknowledgment:} part of the work presented in this paper was funded by the Hochschulraumstrukturmittel (HRSM) provided by the Austrian Ministry
for Research (bmwfw).}

\begin{discussion}
\end{discussion}

\end{document}